\documentclass[aps,pre,twocolumn,groupedaddress]{revtex4}
\usepackage{amsmath,amssymb,amsfonts,graphicx}
\begin{document}

\title{Distance weighted city growth}

\author{Diego Rybski}
\email[]{ca-dr@rybski.de}
\affiliation{Potsdam Institute for Climate Impact Research -- 
14412 Potsdam, Germany, EU}

\author{Anselmo Garc{\'i}a Cant{\'u} Ros}
\affiliation{Potsdam Institute for Climate Impact Research -- 
14412 Potsdam, Germany, EU}

\author{J{\"u}rgen P. Kropp}
\affiliation{Potsdam Institute for Climate Impact Research -- 
14412 Potsdam, Germany, EU}
\affiliation{
University of Potsdam, Dept. of Geo- \& Environmental Sciences, 
14476 Potsdam, Germany}

\date{\today, \jobname}

\begin{abstract}
Urban agglomerations exhibit complex emergent features of which Zipf's law, 
i.e.\ a power-law size distribution, and fractality may be regarded as the most
prominent ones. 
We propose a simplistic model for the generation of city-like structures which 
is solely based on the assumption that growth is more likely to take place 
close to inhabited space. 
The model involves one parameter which is an exponent determining how strongly 
the attraction decays with the distance. 
In addition, the model is run iteratively so that existing clusters 
can grow (together) and new ones can emerge.
The model is capable of reproducing the size distribution 
and the fractality of the boundary of the largest cluster. 
While the power-law distribution depends on both, the imposed exponent and the
iteration, the fractality seems to be independent of the former and only 
depends on the latter. 
Analyzing land-cover data we estimate the parameter-value 
$\gamma\approx 2.5$ for Paris and it's surroundings.
\end{abstract}


\maketitle

\section{Introduction}

Cities or urban agglomerations exhibit signatures of complex phenomena, such
as broad size distributions 
\cite{AuerbachF1913,ZipfGK1949,SaichevMS2010,RozenfeldRGM2011,BerryOK2012} and
fractal structure \cite{BattyL1994,BattyL1987} (and references therein).
The last decades have witnessed a strong interest within the scientific
community in characterizing the worldwide urbanization phenomenon. 
This line of research has strongly benefited from accessibility of 
demographic databases and from application of tools originated in 
statistical physics enabling the identification and analysis of 
universal aspects of urban forms and scaling features \cite{BettencourtW2010}. 
Beyond the descriptive level, various attempts to obtain insights into 
mechanisms that underly the complex features of cities have been proposed.

(i) Multiplicative models 
\cite{LevyS1996,GabaixX1999,MalcaiBS1999,GabaixX2009} 
have explored the connection between random city growth and 
city size distributions. 
In particular, building on discrete random walk theory, 
multiplicative models have proved successful at reproducing Zipf's law 
(i.e.\ power-law city size distribution with exponent close to $2$).
Furthermore, some of these models have proposed plausible explanations for the
origin of these mechanisms, based on spatial economics theory
\cite{GabaixX1999}.
Notwithstanding this fact, multiplicative models are
space-independent and thus are  unable to address other important features of
city structures, such as self-similarity. 
(ii) Approaches based on cellular automata have been used to model 
spatial structure of urban land use over time \cite{WhiteE1993} 
reproducing fractal properties.
(iii) The correlated percolation model (CPM)
\cite{MakseHS1995,MakseABHS1998} assumes that an urban built 
environment is shaped
by spatial correlations, where the occupation probabilities of two sites are
more similar the closer they are. 
The model involves the empirical findings on the radial decay of 
density around a city center.
For certain ranges in the space of parameters, the CPM reproduces basic features
of real urban aggregates, such as broad size distributions in urban clusters and
the fractal scaling of the perimeter.
(iv) Reaction diffusion models \cite{ZanetteM1997,MarsiliMZ1998c,ZanetteM1998r}
have been introduced in order to explore the role of intermittency in creating
spatial inhomogeneities, in agreement with Zipf's law.
(v) Spatial explicit preferential attachment has been shown to be capable
of reproducing Zipf's law \cite{SchweitzerS1998}. 
Here, the probability that a city grows is essentially assumed to be 
proportional to the size of the city.
(vi) Agent based modeling has been employed to simulate urban growth 
\cite{SchweitzerF2003}, reproducing the formation of new clusters as well 
as the merging of neighboring ones.
(vii) A random group formation is presented in \cite{BaekBM2011} from 
which a Bayesian estimate is obtained based on minimal information. 
It represents a general approach for power-law distributed group sizes.

While the term \emph{demographic gravitation} 
was coined by \cite{StewartJQ1948}, in
geographical economics, gravitational models have been
investigated for many decades. \citet{CarrothersGAP1956} provides a review of
gravity and potential concepts of human interaction. 
The so-called Reilly's law of retail gravitation describes the 
breaking point of the boundary of equal attraction \cite{ReillyWJ1931}.
Similarly, Huff's law of shopper attraction \cite{HuffDL1963} provides the
probability of an agent at a given site to travel to a particular facility. 
Last but not least, the volume of trade between countries has been described 
from the point of view of gravity analogy \cite{PoyhonenP1963}.
In contrast, limitation of gravitational models have been pointed out 
in the context of mobility and migration \cite{SiminiGMB2012}.

Following the first law of geography "Everything is related to everything else,
but near things are more related than distant things" \cite{ToblerWR1970}, 
we elaborate on the role of gravity effects in shaping the most
salient universal features of cities, namely, size distribution and fractality.
To this end, we introduce a model where individual lattice sites of a grid 
are more likely to be occupied the closer they are to already occupied sites. 
We find that the cluster sizes follow Zipf's law 
except for the largest cluster which out-grows Zipf's law, 
i.e.\ the largest cluster is too big and can be considered as 
Central Business District \cite{MakseHS1995}.
Applying box-counting \cite{BundeH1994}, we find self-similarity of 
the largest cluster boundary whereas the fractal exponent seems to be 
independent of the chosen exponent.
Despite being very simple, our model intrinsically features radial
symmetry, as in (ii), and preferential attachment, as in (iv).

\section{Model}

We consider a two dimensional square lattice of size $N\times N$
whose sites $w_j$ with coordinates $j\in\{(1\dots N,1\dots N)\}$ 
can either be empty or occupied. 
We start with an empty grid ($w_j=0$ for all $j$) and,
without loss of generality, 
set the single central site as occupied 
($w_j=1$, $j=(N/2,N/2)$ for even $N$, $j=((N+1)/2,(N+1)/2)$ for odd $N$). 
Then the probability that the sites will be occupied is
\begin{equation}
q_j=C\frac{\sum_{k\ne j}w_{k}d_{j,k}^{-\gamma}}{\sum_{k\ne j}d_{j,k}^{-\gamma}}
\enspace ,
\label{eq:model}
\end{equation}
where $d_{j,k}$ is the Euclidean distance between the sites $j$ and $k$. The
proportionality constant $C$ is determined by normalization,
i.e.\ $C=1/\text{max}(q_j)$, so that the maximum probability is $1$. 
The exponent~$\gamma>0$ is a free parameter that determines how strong 
the influence of occupied sites decays with the distance.
This model is inspired by Ref.~\cite{HuffDL1963}, where the probability, 
that a site will be occupied, is solely determined by the distance to 
already occupied sites.

It is apparent that only sites within close proximity of the initially 
occupied site are likely to be occupied, while distant sites mostly 
remain empty.
The procedure is then iterated by repeating the process, 
involving recalculation of Eq.~(\ref{eq:model}) for each step. 
Note that a different choice of $C$ would only influence how many 
iterations are needed to completely fill the lattice.

\begin{figure*}
\begin{center}
\includegraphics[width=0.7\textwidth]{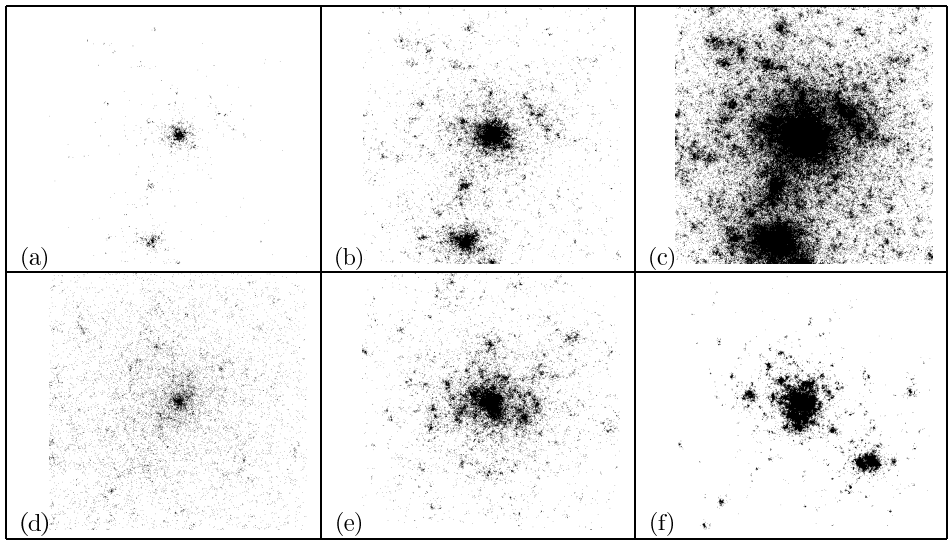}
\end{center}
\caption{\label{fig:example}
Illustrative examples of model realizations for different 
iterations~$i$ and different exponents~$\gamma$.
(a-c) Different iterations of the model $i=6,10,14$ ($\gamma=2.5$, $N=630$). 
Growth takes place close to occupied sites.
(d-f) Realizations with different exponents $\gamma=2.0, 2.5, 3.0$ 
(the occupation probability is $p\approx0.04$, $N=630$).
The smaller~$\gamma$, the more scattered are the emerging structures, 
the larger~$\gamma$, the more compact are they.
}
\end{figure*}

\section{Analysis}

The model output depends on a set of factors. 
Beyond the exponent~$\gamma$, the system size $N\times N$ needs to be chosen. 
As the model works iteratively, the emerging structures can be investigated 
at different iterations~$i$. 
Moreover, we run the model for $M$~realizations in order to obtain better 
statistics.

Figure~\ref{fig:example} shows examples of model realizations. 
Visually, the emerging structures could be associated with urban space.  
Figure~\ref{fig:example}(a-c) shows three iterations of a single realization. 
For high values of $\gamma$ the spatio-temporal evolution is strongly 
influenced by the sites which are occupied early, 
see Figure~\ref{fig:example}(d-f). 
Such a path dependency is also reflected in the reduction of rotational 
symmetry observed for increasing values of $\gamma$. 
In particular, the larger~$\gamma$ is chosen, the more compact and 
less scattered are the obtained structures. 
Large~$\gamma$ also leads to slower filling of the lattice.

\subsection{Cluster size distribution}

\begin{figure}
\begin{center}
\includegraphics[width=0.8\columnwidth]{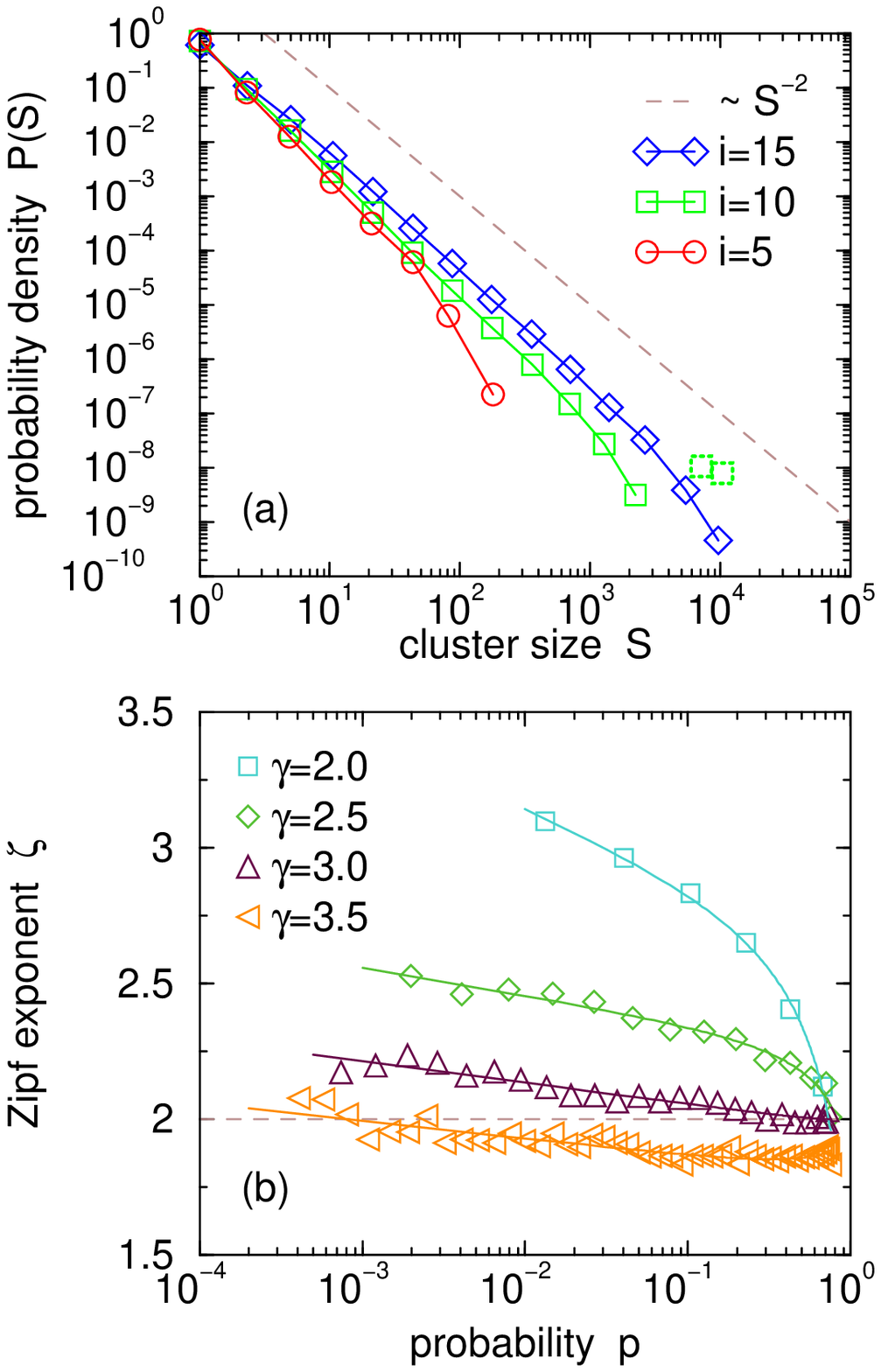}
\end{center}
\caption{\label{fig:zipf1}
(Color online)
Cluster size distribution and dependence of the Zipf exponent on the 
occupation probability.
(a) Examples of probability density distributions~$P(S)$ 
of the cluster size~$S$ \emph{disregarding the largest cluster} 
of each realization: 
$i=5,10,15$ (from top to bottom, $p\approx 0.004, 0.078, 0.573$), 
$\gamma=2.5$, $N=630$, $M=100$ (solid lines connect symbols). 
The two green dotted squares represent the contribution of the largest cluster
($i=10$).
(b) The obtained Zipf exponents $\zeta$ as a function of the 
occupation probability~$p$: 
$\gamma=2,2.5,3,3.5$ (from top to bottom), $N=630$, $M=100$. 
Solid lines are fits according to Eq.~(\ref{eq:zeta}). 
Dashed lines indicate $\zeta=2$.
}
\end{figure}

We begin our analysis by studying the cluster size distribution. 
We employ the City Clustering Algorithm (CCA) 
\cite{RozenfeldRABSM2008,RozenfeldRGM2011} and find that the largest cluster 
is markedly larger than the remaining ones (Fig.~\ref{fig:zipf1}(a)), 
i.e.\ larger than expected from Zipf's law.
The presence of such anomalous extremes in size distributions are denoted as 
\emph{Dragon Kings} and are signatures of strongly cooperative dynamics 
\cite{PisarenkoS2011}. 
A similar effect has been found in another model \cite{SchweitzerS1998}, 
where -- in order to avoid their appearance -- the domination of the 
largest cluster is inhibited by excluding it from proportionate growth. 
Exclusion is not feasible in our model and we omit it when studying 
the cluster size distribution \cite{MakseHS1995,MakseABHS1998}.

Figure~\ref{fig:zipf1}(a) shows examples of the probability density~$P(S)$ of 
the cluster size~$S$ disregarding the largest cluster of each realization. 
We find approximate power-laws according to
\begin{equation}
P(S)\sim S^{-\zeta}
\, ,
\label{eq:zipf}
\end{equation}
where $\zeta$ is the Zipf exponent.
In Fig.~\ref{fig:zipf1}(a) one can see, deviations from Eq.~(\ref{eq:zipf}), 
in the form of too few large clusters.
Naturally, for late iterations Eq.~(\ref{eq:zipf}) extends over more 
decades of cluster size than for early iterations.
As can be seen, the Zipf exponent is close to $2$. 
To be more precise, $\zeta$ is smaller for large iteration~$i$ 
than for small~$i$.
Accepting minor deviations from $\zeta=2$, the model produces 
cluster size distributions compatible with Zipf's law.

In order to better understand how $\zeta$ relates to the model parameters, 
we express the iteration~$i$ in terms of the 
overall occupation probability~$p$ which for a given $i$ is 
defined by the number of occupied sites divided by the 
total number of sites, i.e.\ $N\times N$. 
The probability~$p$ increases monotonically with the iteration~$i$.
In Fig.~\ref{fig:zipf1}(b) $\zeta$ is plotted as a function of $p$. 
As can be seen, it decreases monotonically with increasing probability and 
strongly depends on the model exponent $\gamma$. 
While for $\gamma<3$, convex $\zeta(p)$ are found with overall $\zeta>2$, 
for $\gamma=3$, an almost logarithmic form can be identified with $\zeta>2$ 
and $\zeta\rightarrow 2$ for $p\rightarrow 1$.
In contrast, for $\gamma>3$, we see a slightly concave relation and $\zeta<2$ 
(except for small $p$).
Accordingly, the Zipf exponent depends strongly on both, 
the model exponent $\gamma$ and the iteration of the model $i$.

\begin{figure}
\begin{center}
\includegraphics[width=0.8\columnwidth]{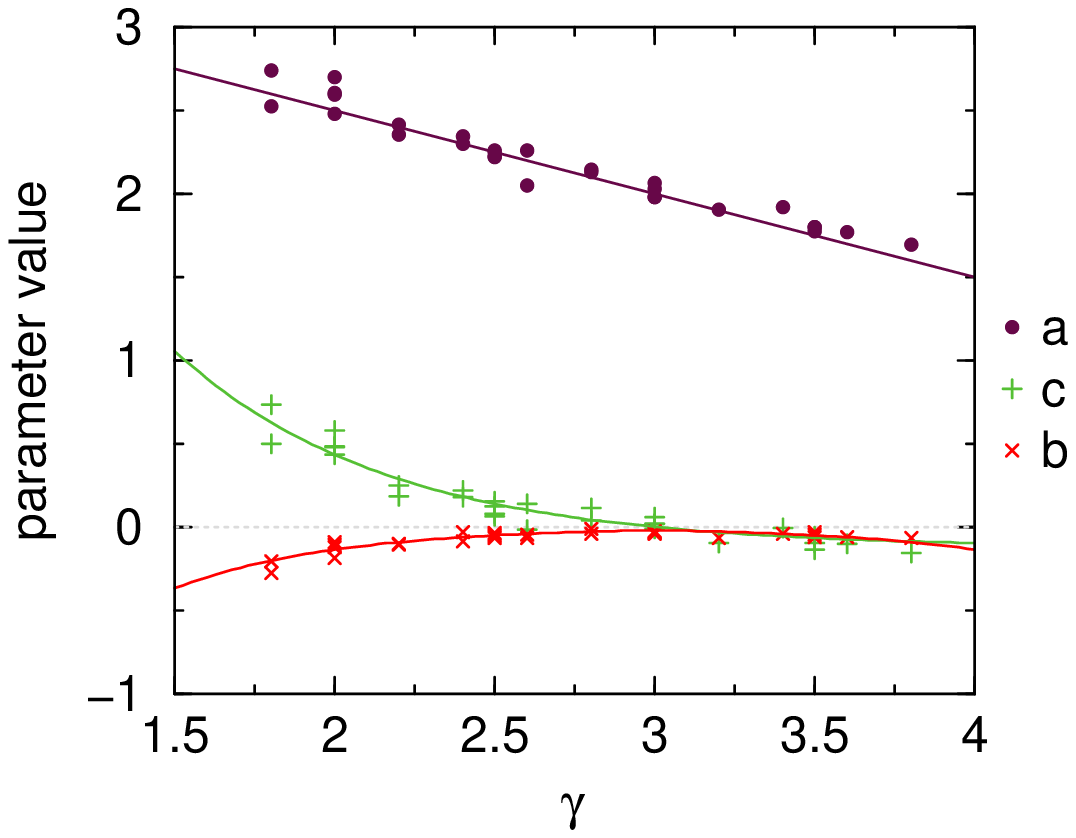}
\end{center}
\caption{\label{fig:zipf2}
(Color online)
Parameter values $a,b,c$ as a function of the exponent $\gamma$ from fitting 
the Zipf exponent versus the occupation probability (Fig.~\ref{fig:zipf1}(b)).
The dots represent the parameters obtained from fitting Eq.(\ref{eq:zeta}) 
to the numerical values, the solid lines are given by 
$a = -k_1\gamma+k_2$, 
$b_{\gamma\le 3} = -{\rm e}^{-k_3\gamma+k_4}$ and 
$b_{\gamma>3} = -{\rm e}^{k_3\gamma-k_5}$, and 
$c = {\rm e}^{-k_6\gamma+k_7}-{\rm e}^{-k_8+k_7}$
($k_1$-$k_8$ are fitting parameters).
In order to have enough values, we do not separate the different system sizes.
}
\end{figure}

Moreover, we find that the dependence of $\zeta$ on $p$ can be well 
approximated by
\begin{equation}
\zeta(p)=a+b\ln(p)+c\ln(1-p) \, ,
\label{eq:zeta}
\end{equation}
which also is the logarithm of a beta-distribution.
The solid lines in Fig.~\ref{fig:zipf1}(b) are non-linear fits to 
the numerical model results, providing the fit parameters $a$, $b$, 
and $c$.
In Fig.~\ref{fig:zipf2} the obtained values of these parameters 
are plotted against $\gamma$.

\subsection{Fractality}
\label{ssec:fractal}

Next we analyze fractal properties of the \emph{urban envelope} 
of the largest cluster. 
Therefore, we first extract the boundary of the cluster. 
This is done, by identifying those largest cluster sites 
which have at least one empty neighboring cell which connects 
to the system border via a nearest neighbor path of empty sites
(the latter condition is necessary to exclude inclusions).
Thus, here the boundary is defined as the occupied neighbors of 
the external perimeter \cite{GrossmanA1987}.
Then we apply box-counting, i.e.\ perform coarse-graining and 
count how many sites or occupied. 
Thus, we regularly group $m\times m$ sites and accordingly reduce 
the system size to $(N\times N)/(m\times m)$. 
Finally, we count the number of occupied sites $N_\text{B}$ 
for a chosen box size~$m$.

\begin{figure}
\begin{center}
\includegraphics[width=0.8\columnwidth]{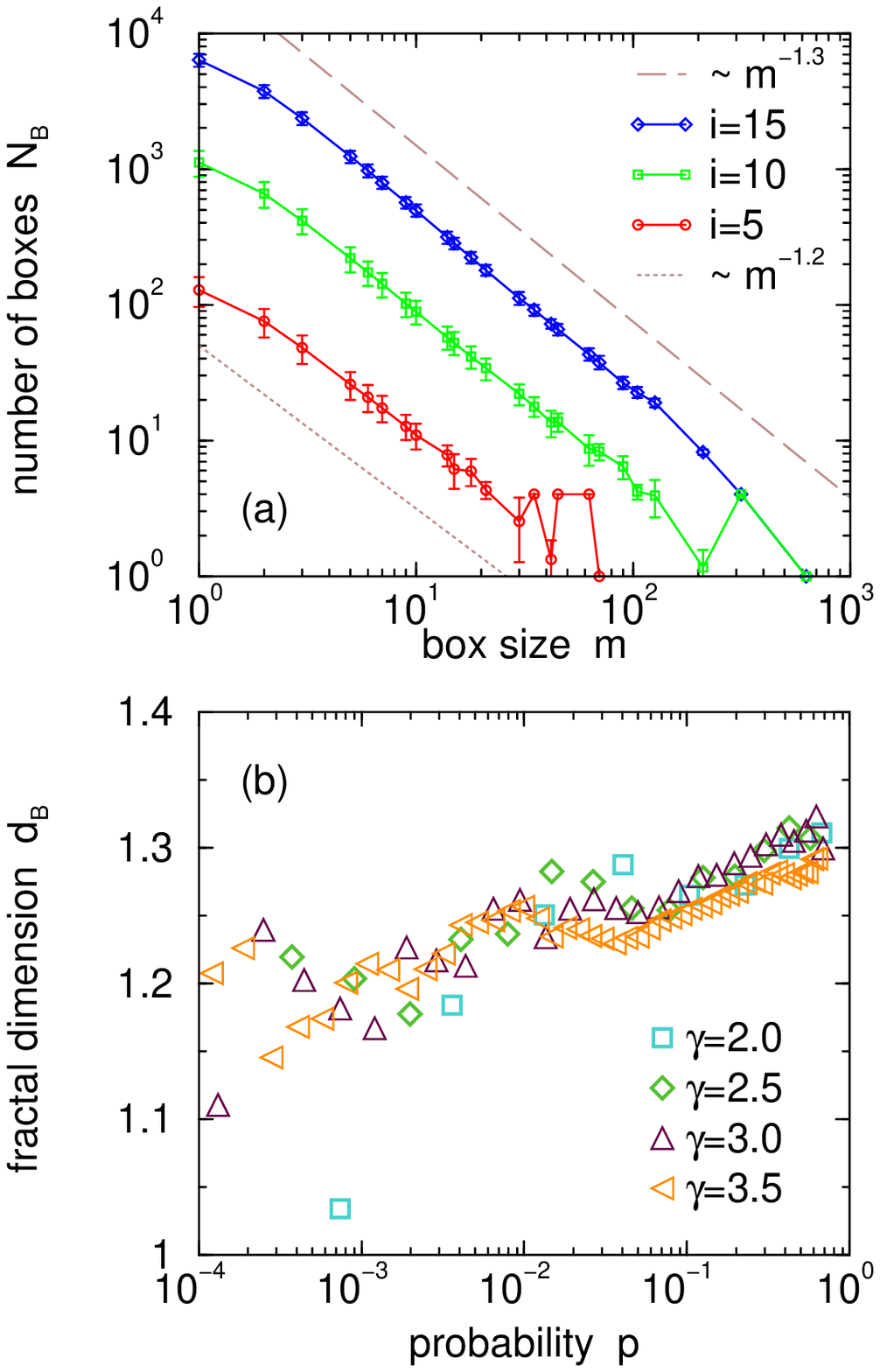}
\end{center}
\caption{\label{fig:fractal}
(Color online)
Self-similar scaling of the boundary of the largest cluster.
(a) Number of boxes necessary to cover the boundary as a function of the size of
the boxes: $\gamma=2.5$, $N=630$, $i=5,10,15$ 
($p\approx 0.004, 0.078, 0.573$), $M=100$. 
The results follow Eq.~(\ref{eq:boxcov}) indicating the fractal property of 
the boundary with a fractal dimension~$1<d_\text{B}<2$.
The error bars represent standard deviations among the realizations.
(b) Fractal dimension as function of the occupation probability for 
$\gamma=2.0,2.5,3.0,3.5$ ($N=630$, $M=100$). 
$d_\text{B}$ increases approximately logarithmically, independent of $\gamma$.
}
\end{figure}

Examples of $N_\text{B}(m)$ are displayed in Fig.~\ref{fig:fractal}(a).
Apart from minor deviations for small and large~$m$, straight lines 
are found in the log-log representation, corresponding to
\begin{equation}
N_\text{B}(m)\sim m^{-d_\text{B}}
\, ,
\label{eq:boxcov}
\end{equation}
where $d_\text{B}$ is a measure of the fractal dimension of the 
cluster boundary. 
In the displayed examples, we approximately 
find $d_\text{B}\approx 1.25$.

Figure~\ref{fig:fractal}(b) shows $d_\text{B}$ as a function of the 
occupation probability~$p$. 
Qualitatively, we find a logarithmic dependence, 
implying $d_\text{B}\rightarrow \text{const.}$~for~$p\rightarrow 1$, 
which seems to be independent of $\gamma$. 
Overall, the values are clearly below those expected from uncorrelated
percolation slightly above or below the 
percolation transition \cite{VossRF1984}. 
This difference could be due to inherent correlations in our model.
Nevertheless, the model generates self-similar (fractal) largest clusters.
The evolving fractal dimension is at least qualitatively consistent with 
urban areas, see e.g.\ \cite{ShenGQ2002}.

Last, we would like to note that the definition of the boundary has 
a substantial influence on the fractal dimension \cite{GrossmanA1987}.
Moreover, box-counting results can differ from those obtained with other 
techniques such as the equipaced polygon method \cite{KayeC1985}.
Further analysis is required to shed light on these aspects.

\subsection{Percolation transition}

\begin{figure}
\begin{center}
\includegraphics[width=0.8\columnwidth]{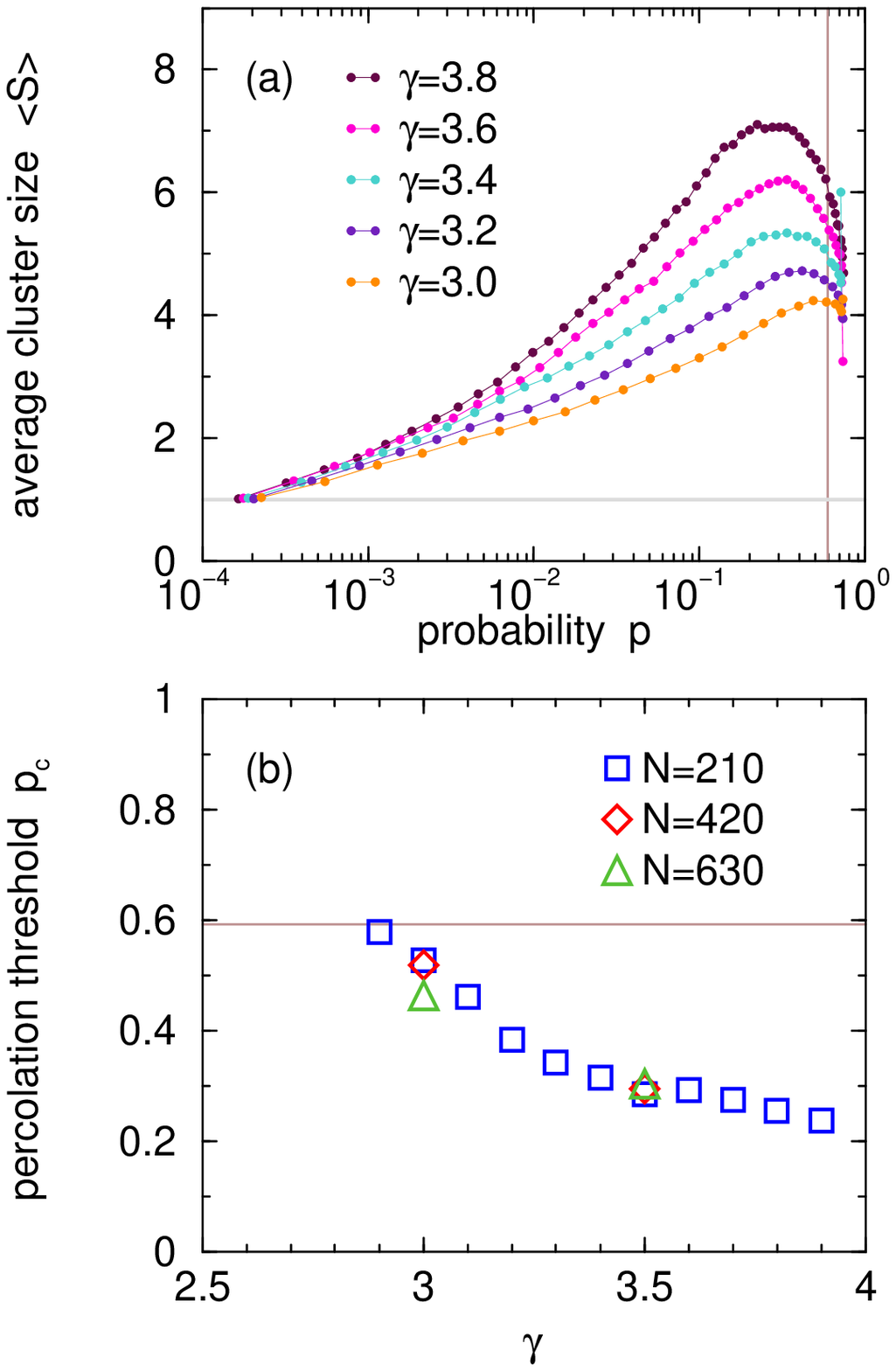}
\end{center}
\caption{\label{fig:transition}
(Color online)
Percolation transition of the model.
(a) Average size of the \emph{finite} clusters: 
$\gamma=3.8, 3.6, 3.4, 3.2, 3.0$ (from top to bottom), $N=210$, $M=1000$. 
The vertical line represents the percolation transition 
of uncorrelated percolation $p^*_\text{c}\simeq 0.593$ 
(site percolation in the square lattice, \cite{BundeH1991}). 
The maximum is located at the percolation transition.
(b) Percolation transition $p_\text{c}$ as a function of $\gamma$: 
$N=210, 420, 630$ ($M=1000, 400, 100$). 
For small $\gamma$ the percolation transition is close to $p^*_\text{c}$ as 
indicated by the horizontal line.
}
\end{figure}

One may argue that at certain iteration the system might undergo a percolation 
transition. 
Thus, finally we characterize the percolation threshold of the model.
Therefore, we calculate the average cluster size disregarding the largest 
component, $\langle S\rangle$, as a function of the occupation probability~$p$. 
At the percolation transition, $p_\text{c}$, the average cluster size 
exhibits a maximum \cite{BundeH1991}. 
Figure~\ref{fig:transition} depicts $\langle S\rangle(p)$ for some values 
of $\gamma$.
A distinct peak can be found which moves to larger~$p$ with decreasing~$\gamma$.
For $\gamma<3$ the maximum becomes less clear and 
we cannot determine $p_\text{c}$.

The obtained percolation thresholds are plotted versus $\gamma$ in 
Figure~\ref{fig:transition}(b). 
The transition decreases monotonically with increasing~$\gamma$. 
For $\gamma\approx 3$ the value is close to the transition 
of uncorrelated site percolation in the square lattice 
($p^*_\text{c}\approx 0.593$, \cite{BundeH1991}).
For $\gamma\approx 4$ we find $p_\text{c}\approx 0.2$.

We would like to note that the results of Zipf and fractality analysis 
seem to be independent from percolation transition, 
i.e.\ there is no change in the behavior below or above $p_\text{c}$.
Accordingly, scaling in the form of Zipf's law and fractality are 
reproduced even away from criticality.

\section{Analyzing real data}

\begin{figure}
\begin{center}
\includegraphics[width=0.9\columnwidth]{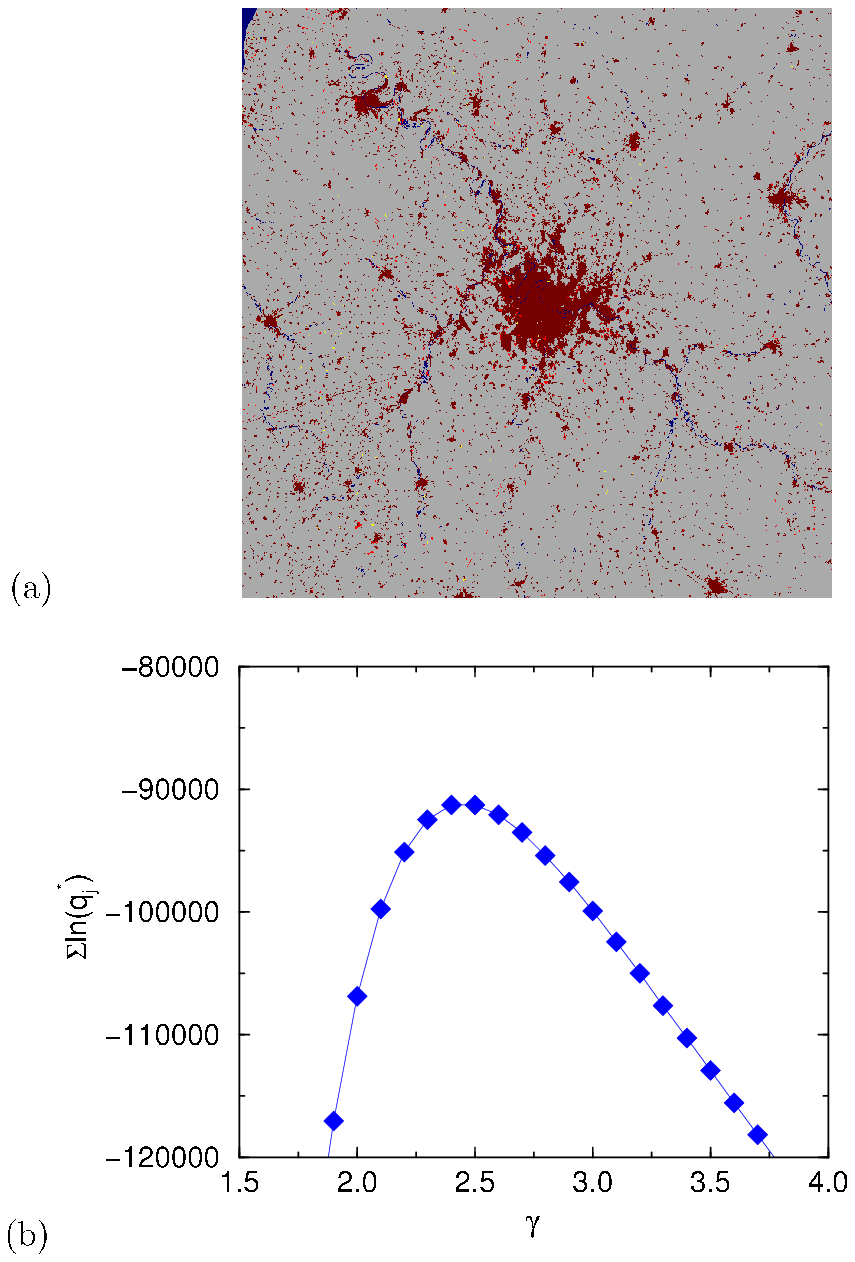}
\end{center}
\caption{\label{fig:paris}
(Color online)
Urban growth and the estimation of $\gamma$ for Paris and it's surroundings 
in the years 2000 and 2006. 
(a) Considered land-cover data in a window of $1000\times 1000$ grid points 
($250$\,m resolution). 
The panel distinguishes the following sites: 
populated in 2000 (dark red), unpopulated in 2000 (grey), water (blue), 
non-urban $\rightarrow$ urban (light red), and 
urban $\rightarrow$ non-urban (yellow).
(b) Log-likelihood of urbanization. 
First we calculate the probabilities $q_j$ according to Eq.~(\ref{eq:model}) 
for the year 2000, whereas ($w_k=1$ for urban and $w_k=0$ for non-urban cells). 
Then we determine $Q$ according to Eq.~(\ref{eq:loglike}) 
over all non-urban cells in 2000 
(the change urban to non-urban is disregarded). 
We find a maximum at $\gamma_\text{Paris}\approx2.5$.
}
\end{figure}

Finally, it is of interest which $\gamma$-value real city growth exhibits. 
In order to address this question, we consider Paris and its surroundings. 
We analyze CORINE \cite{BossardFO2000} land-cover data in $250$\,m 
resolution and only distinguish between urban and non-urban land grid cells. 
For the years 2000 and 2006 we extract a window of 
$1000\times 1000$ grid points (Fig.~\ref{fig:paris}(a)) and study the 
land-cover change. 
Since our model only includes growth, we focus on those cells which 
change from non-urban to urban and disregard the opposite.

First we calculate the probabilities $q_j$ according to Eq.~(\ref{eq:model}) 
for the year 2000, with $w_k=1$ for urban and $w_k=0$ 
for non-urban cells. 
Then we determine the log-likelihood by summing over all 
non-urban cells in 2000,
\begin{equation}
Q= \sum_{j\in A}\ln q_j + \sum_{j\in B}\ln (1-q_j)
\, , 
\label{eq:loglike}
\end{equation}
where $A$ is the set of cells changing from non-urban to urban and 
$B$ is the set of cells remaining non-urban.
Varying $\gamma$ we can identify the value for which $Q$ is maximized, 
i.e.\ for which the probabilities calculated with Eq.~(\ref{eq:model}) 
best represent the non-urban to urban land-cover change in the real data. 
As can be seen in Fig.~\ref{fig:paris}(b), the maximum is located at 
$\gamma_\text{Paris}\approx2.5$. 
The qualitative similarity between 
Fig.~\ref{fig:paris}(a) and Fig.~\ref{fig:example}(e) 
supports this quantitative result, but the comparison also shows 
that the real example is richer in structure. 
While the analysis does not provide sufficient evidence to support our model, 
it leads to the value of $\gamma$ for which the model best fits the growth of 
Paris.

\section{Discussion}

We also find that the growth rate of clusters between two iterations is
independent of the cluster size (not shown). 
This implies proportionate growth, a characteristic which is also 
featured by preferential attachment \cite{SimonHA1955,BarabasiA1999}. 
We would like to note that in the proposed model such mechanism emerges 
and is not included explicitely. 
We further find that the standard deviation of the growth rate
decays as a power-law with exponent $1/2$ (not shown) which indicates
uncorrelated growth \cite{RozenfeldRABSM2008,RybskiBHLM2009}. 
Again, analyzing the growth, we have disregarded the largest cluster.

While the work in hand briefly introduces our model, 
more research is necessary to characterize it. 
This includes
(i) an analytical description of the model, 
(ii) further numerical analysis, in particular refining the 
fractal characterization (as mentioned in Sec.~\ref{ssec:fractal}) 
or other features such as the area-perimeter scaling \cite{AnderssonLRW2002}, 
and 
(iii) relating our model to other physical approaches, 
such as \cite{AnderssonLRW2002,JeonC2005,ReeS2006}.

The analogy of gravitation has a long history in geography and spatial economy.
However, the early works were limited by scientific background from statistical
physics as well as computational power.
Here we reexamine the concept of \emph{gravity cities} by proposing a simple
statistical model which generates city-like structures. 
The emergent complex structures are similar to urban space.
On the one hand, we find that the largest cluster which can be considered 
as Central Business District \cite{MakseHS1995} exhibits fractality, 
consistent with measured urban area. 
On the other hand, clusters around the largest one 
can be considered as towns surrounding a large city \cite{MakseHS1995}. 
Their cluster size distribution is compatible with Zipf's law.

\begin{acknowledgments}
We would like to thank B.F. Prahl, B. Zhou, L. Kaack, E. Giese, X. Gabaix, 
H.D. Rozenfeld, and N. Schwarz for useful discussions and comments. 
We appreciate financial support by BaltCICA 
(part-financed by the EU Baltic Sea Region Programme 2007-2013).
\end{acknowledgments}


%

\end{document}